\documentclass[11pt,a4paper]{article}
    \usepackage{jheppub}
    \usepackage[T1]{fontenc}     
    
   \usepackage{subcaption}
   \usepackage{float}
   \usepackage{array}

\DeclareMathOperator{\Tr}{Tr}

\title{Baryons in $SO(N)$ vector models \\and their duals in higher spin theory}

\author{Bo Sundborg}

\affiliation{The Oscar Klein Centre \& Department of Physics, Stockholm University,\\
AlbaNova, 106 91 Stockholm, Sweden}

\emailAdd{bo.sundborg@fysik.su.se}

\abstract{
Black shells, a kind of black hole mimickers, are identified thermodynamically as bulk duals of baryon operators in vector models, indicating that such objects are essential for the consistency of higher spin gravity theories. Thermal baryons, with a spectrum of a 2+1-dimensional relativistic Fermi gas, are found to be precursors of the deconfinement phase transition in vector models, condensing at a slightly lower temperature. The early condensation means that baryons are statistically important already in the phase with weakly interacting higher spin fields. Furthermore, the mysterious scale of the deconfinement transition in vector models is naturally interpreted as the Fermi energy scale in the gas. 
}

\begin{document}

\maketitle
\flushbottom

\section{Introduction}
Large $N$ conformal field theories generally have both light and heavy operators, where the light operators are dual to standard bulk particles and heavy operators are dual to very massive objects, possibly black hole microstates. 
Concrete descriptions of heavy states should exemplify how large $N$ CFTs encode gravitational phenomena, and generic mechanisms are likely to be clearer in simple models. To test this concept, we consider super-heavy baryon operators in free vector models, which are simpler than matrix model operators and have a rich but explicitly known spectrum, which we review. We are led to explore the thermodynamics of these states and find a match to the thermodynamics of exotic bulk objects, as well as signs of a new condensation phenomenon slightly below the critical deconfinement temperature of the model. 

The spectra of heavy operators in holographic theories at finite $N$ is connected to black hole spectra. It is expected from early days that the characteristically large (but finite) entropy of black holes should be related to finite $N$ relations \cite{Susskind:1994sm,Witten1998}, a fact also manifested in free boundary theories \cite{Sundborg2000}. The present study provides a direct link between a heavy spectrum controlled by finite $N$, bulk objects and thermodynamics. Although the field is mathematically and technically challenging, there has recently been substantial insights and developments in several directions, supersymmetric black hole counting \cite{Cabo-Bizet:2018ehj,Choi:2018hmj}, finite $N$  fortuity and supersymmetry \cite{Chang:2022mjp,Chang:2024zqi} as well as trace relations at finite $N$ \cite{deMelloKoch:2025ngs}. The example of vector model baryons is simpler than many other cases, and our findings are likely to be reflected in more complex settings.

Vector models are related to bulk theories of massless higher spin fields \cite{KlebanovPolyakov2002}. There are different approaches to the bulk theory, the original \cite{Vasiliev:1990en,Vasiliev1999}, and the collective approach \cite{KochJevickiJinRodrigues2011,Mello-KochJevickiSuzukiYoon2019,Aharony:2020omh} which more or less constructs the bulk theory from the boundary theory. In ref.\ \cite{Aharony:2022feg} a framework which accommodates finite $N$ and thermal effects has also been developed. Unlike string theory, higher spin theory has not been related to an extended object and its symmetries. The example of string theory indicates that such a connection can be pivotal, and it is especially promising for higher spin theory due to its previous relation to string theory, by appearing in its tensionless limit \cite{Sundborg2001}. The  heavy baryon operators \cite{ShenkerYin2011} in the boundary theory are richer candidates than light operators for describing extended objects, and we hope the bulk picture of them that we establish in this work will be fruitful.


The paper is organised as follows. In section \ref{sect_SimpleStates} we review the operator content of $SO(N)$ vector models and baryon operators in particular. It is followed by the statistical mechanics study of the baryon spectrum and sub-deconfinement condensation in section \ref{sec_BoundaryBaryon}, and the study of potentially dual bulk objects in section \ref{sect_DualObjects}. Boundary and bulk objects are matched in section \ref{sect_Discussion} and section \ref{sect_Conclusion} presents our conclusion.

\section{Simple states in vector models}\label{sect_SimpleStates}
We consider free vector models on $S^2\times R$, and will eventually focus on vectors of scalar fields transforming in the fundamental representation of $SO(N)$ though some of the discussion will appeal to results from $U(N)$ and $SU(N)$ symmetry. In those cases the field content is instead one fundamental representation and one anti-fundamental representation. We will be interested in invariants of all these transformation groups, with the understanding that physical operators and states are group invariants. From the perspective of the scalar fields the invariance requirement may be formalised by gauging the symmetry and taking the coupling to zero. Alternatively, we may consider the closed operator product algebra of invariant operators as the definition of the CFT. In either case, the spectrum of local operators or by CFT operator-state duality, the spectrum of states on $S^2$, is of central importance. 

\subsection{Simple light operators in vector models}

Single particle states in the bulk are typically represented by single trace operators in the boundary theory. They are not our main focus but important, and we describe them here also to illustrate the sharp distinction between light and heavy operators in vector models. In matrix models single trace operators are actual traces, eg $\Tr \Phi^2$, and are contrasted with double trace operators, for example $\Tr \Phi^2\Tr \Phi^2$, while for the present case, vector models, the term single trace is often used for invariants with a single contraction of vector indices, as in $\phi^\dagger \phi$. Then $\phi^\dagger  \phi \,\phi^\dagger \phi$
 is an example of a double trace operator. 
 
 Derivatives on one of the constituent fields in a single trace operator generally produces a current, such as $\phi^\dagger \partial_\mu\phi$. For counting single trace operators including their descendants (derivatives) we would double count the internal derivative configurations, by Leibniz rule, if we counted derivatives acting on more than one constituent. Thus we count derivatives on one field, meaning that we should count symmetric spacetime tensors. However, $\Box=\eta^{\mu\nu}\partial_\mu\partial_\nu$ acting on a massless free field vanishes and such operators do not count, effectively leaving us with conserved and symmetric traceless currents. 
Furthermore, the transverse part of each tensor is related to a descendant of a lower tensor and does not represent new degrees of freedom. All single trace operators are thus conserved higher spin currents, which correspond to massless higher spin fields in the bulk \cite{Flato:1980zk}.
 
 A crucial property of the basis of single trace operators is that the basis elements 
 do not factorise into other invariant operators (like multi-trace operators do). There are however other invariant operators which cannot be written in terms of single trace operators. They are the main topic of this work.

\subsection{Simple heavy operators in vector models}\label{HeavyOperators}

 We now turn to a more interesting class of simple invariant operators, which are called `baryonic' due to anti-symmetrisation of the constituent fields in the operators, generalising the total anti-symmetrisation of the fundamental representations of $SU(3)$ quarks in a baryon, see refs. \cite{ShenkerYin2011,Aharony2016}. There are more general anti-symmetrised operators which also belong to a class of operators called \emph{secondary invariants} \cite{deMelloKoch:2025ngs}.\footnote{The role of anti-symmetrisations is clear for secondary invariants in vector models \cite{deMelloKoch:2025cec}.}  Secondary invariants are more general and in many ways behave as baryons, but baryons are simpler to deal with, by always having a description independent of the choice of basis in the space of operators. Baryons are our object of study, but we note that there are typically many more operators with similar properties, also in models without baryons, eg.\ the $U(N)$ vector model.

 For $SU(N)$, in the presence of totally antisymmetric Levi-Civita symbols, invariant operators of the form 
\begin{equation}\label{eq:HeavySO(N)Operator}
H= \epsilon^{a_1\dots a_N}\partial^{l_1}\phi_{a_1}\partial^{l_2} \phi_{a_2} \dots \partial^{l_N} \phi_{a_N}, \quad \text{or} 
\quad \bar{H}= \epsilon_{a_1\dots a_N}\partial^{l_1}\bar{\phi}^{a_1}\partial^{l_2} \bar{\phi}^{a_2} \dots \partial^{l_N} \bar{\phi}^{a_N},
\end{equation}
are constructed from contractions with a product of $N$ constituent field operators differentiated $l_i$ times \cite{ShenkerYin2011}. Here $l_1 \le l_2 \le \dots \le l_N$, and the inequalities just represents a way of ordering the factors in order to avoid double counting. Different differential operators acting on the different factors are of course crucial for the operators to be non-vanishing. The differential operators themselves produce symmetric traceless indices, just as for the currents in the discussion of light operators above. 

A peculiar and important property of baryon operators is their behaviour under products. We shall study the $SO(N)$ case, which is even simpler than $SU(N)$. Then, the fields $\phi_a$ are real and there is no distinction of upper and lower indices. Consequently, there is no distinction between $H$ and $\bar{H}$ operators. That the product of two Levi-Civita symbols is a totally antisymmetrised product of Kronecker deltas dissolves operators of the form $H_1 H_2$ into a superposition of products of $N$ bilinear operators. Baryons therefore do not behave as bilinears or single trace operators, for which any monomial represents a new independent operator in the large $N$ limit. In contrast to products of bilinear invariants, the products of two $SO(N)$ baryons never count as independent new operators. In the $SO(N)$ vector model, there is thus only one set of single-baryon operators $H$, and no multi-baryon operators.

The spectrum of all single baryon operators can now be described. In $d=3$, traceless symmetric tensors imply $2l+1$ different nontrivial differential operators of order $l$, corresponding to the different states of angular momentum $l$. The full  baryonic $H$ can thus be viewed as a collection of $N$ particles of \emph{different} angular momentum quantum numbers $(l_i,m_i)$. The dimension $\Delta_H$ of the operator $H$ is simply the sum of $N/2$ from the constituent fields and the $l_i$ from the derivatives, since the CFT dual to HS theory is a free vector model. For example, an operator with $N=L^2$ different symmetric derivative operators then has the minimal dimension and energy
\begin{equation}\label{eq:groundstate}
\Delta_0 = N/2+3 \times 1 +\dots +(2L-1)\times (L-1)= \frac{1}{6}L(4L^2-1)\sim \frac{2}{3}N^{3/2}, \quad E_0 \sim \frac{2N^{3/2}}{3R}
\end{equation}
from filling all the smallest terms consistent with the requirement that all differential operators/angular momentum states are different. Here, $R$ is a length scale giving the energy the right dimensions. Thus, the total antisymmetrisation by the $\epsilon$ symbol leads to an effectively fermionic picture of the spectrum as that of $N$ fermions, each with $2l+1$ single particle states at level $l$.\footnote{In this fermion analogy we focus on multiparticle statistics and disregard the absence of spin degrees of freedom.}

The baryon spectrum is actually that of a Fermi gas of $N$ particles in a very concrete sense. By the operator/state correspondence the spectrum of operator dimensions may alternatively be obtained form the energy spectrum of corresponding states on $S^2 \times \mathbb{R}$, built from $N$ massless conformally coupled scalars with Fermi statistics. The conformal coupling gives the energy $E=(l+\frac{1}{2})/R$ to a particle on the sphere of angular momentum $l$ where $R$ is now identified as a measure of the radius of the $S^2$. For the examples above of a perfectly square $N=L^2$, the groundstates are non-degenerate and spinless with filled shells in the shell structure due to the Pauli exclusion principle. In contrast, for general $N$, already the groundstates are highly degenerate. 

The concrete description of the spectrum of  baryonic states in terms of a relativistic gas on $S^2$ is a first hint that they may describe excitable spherical membranes. Fermi gases are characterised well by Fermi momentum, but on a sphere a `Fermi angular momentum', here  $l_F=\sqrt{N}$, is more appropriate, and we will see the importance of the Fermi energy $E_F = l_F/R$ below in the thermodynamics of the baryon spectrum. 

Since the heavy operators $H$ form an $N$ particle system, a large $N$ yields an enormous number of operators, differing in numbers and distributions of derivatives on the factors in the product \eqref{eq:HeavySO(N)Operator} suggesting that a statistical description is useful. The heavy operators clearly outnumber the bilinear higher spin currents schematically written
\begin{equation}
L_s = \sum_{s_1+s_2=s} c^{(s)}_{s_1s_2}\partial^{s_1}\phi_a \partial^{s_2}\phi_a,
\end{equation}
which are dual to bulk massless higher spin fields, but their role is mysterious.

\section{Boundary baryon thermodynamics}\label{sec_BoundaryBaryon}

We now wish to understand a typical baryon operator. It is best described statistically by imagining that the corresponding state is kept in thermal equilibrium with a heat bath at temperature $T$. For many-particle systems of $N$ fermions, the grand canonical partition function $\mathcal Z(\mu, \beta=1/T)$ summing over different $N$ is convenient\footnote{Averaging over $N$, or over theories, has been considered in holography, at times elevated to a principle. This is intriguing, but our motivation is primarily convenience. The shell structure of the state space causes oscillations controlled by simple number theory requiring a certain averaging to secure a smooth description.}, and we tune a chemical potential $\mu(T,N)$ to obtain the expectation value of the fluctuating numbers $n$ of fermions
\begin{equation}\label{eq_ChemPot}
\left< n\right>= T\partial_\mu \log \mathcal Z(\mu,\beta)\equiv N .
\end{equation}
Since we wish to understand typical properties of this system, it is enough if relative fluctuations in $N$ are small, ie $\delta N^2 = \left<n^2\right> - \left< n\right>^2 \ll  N
^2$, which holds in all our final results.

The grand canonical partition function of the states described above is
\begin{equation}
\mathcal Z(x,y,0)\approx\exp \left(-\sum_{m=1}^\infty \frac{(-1)^m}{m} \frac{y^mx^{m/2} (x^m+1) }{(1-x^m)^2}\right)
\end{equation}
in terms of $y=e^{\beta\mu}$ and $x=e^{-\beta/R}$. It can be read off from generating functions in \cite{ShenkerYin2011} which summarise our description of the spectrum in section (\ref{HeavyOperators}). The grand canonical potential 
\begin{equation}
\Phi = U-TS-\mu N=-\frac{1}{\beta}\log \mathcal Z\approx 2\sum_{m=1}^\infty  \frac{(-y)^mR^2 }{  m^3\beta^3 }=\frac{2R^2}{\beta^3}\mathrm{Li}_3(-e^{\beta\mu})=\frac{2N}{r} t^3\mathrm{Li}_3(-e^{\beta\mu})\label{eq:GCP}
\end{equation}
for $\beta \ll R$, ie for high temperature, where we have introduced the natural length scale $r$ and the dimensionless temperature $t$ given by
\begin{equation}
r=\frac{R}{\sqrt{N}},  \qquad t= r T .
\end{equation}
We can verify that the scale $r$ is natural by inspecting the condition \eqref{eq_ChemPot} fixing the chemical potential, or equivalently the fugacity $y$. It reads
\begin{equation}\label{eq:ty}
-2\mathrm{Li}_2(-y)=\frac{1}{t^2}=\frac{\beta^2}{r^2}= N\frac{\beta^2}{R^2}\equiv \frac{E_F^2}{T^2},
\end{equation}
which 
shows the central role of the Fermi energy $E_F$ as a scale for the high temperature limit. The characteristic length scale $r$, which is important for holography, is actually the Fermi wavelength. 

As in eq. \eqref{eq:GCP} for the grand canonical potential, \emph{the extensive thermodynamic functions are  $N$ times a dimensionless function of $t$} with dimension from $r$. It is cumbersome to invert the dilogarithm \eqref{eq:ty}, but it is doable numerically and in some limits, 
\begin{eqnarray}
y =  e^{1/t}\quad &\mathrm{for}&\quad t\to 0,\\
y\to 1\quad &\mathrm{for}&\quad t \to \frac{\sqrt{6}}{\pi},\\
y = 1/2t^2\quad &\mathrm{for}&\quad t\to \infty,
\end{eqnarray}
using
\begin{eqnarray}
\mathrm{Li}_2(-y)&\to& -y\ \mathrm{as}\ y\to 0,\\
\mathrm{Li}_2(-y)&\to& -\frac{\pi^2}{12}\ \mathrm{as}\ y\to 1,\\
\mathrm{Li}_2(-y)&\to& -\frac{1}{2}(\log y)^2 -\frac{\pi^2}{6}\ \mathrm{as}\ y\to \infty.
\end{eqnarray}
We can also make use of
\begin{eqnarray}
\mathrm{Li}_3(-y)&\to& -y\ \mathrm{as}\ y\to 0,\\
\mathrm{Li}_3(-y)&\to& -\frac{3}{4}\zeta(3)\ \mathrm{as}\ y\to 1,\\
\mathrm{Li}_3(-y)&\to& -\frac{1}{6}(\log y)^3 -\frac{\pi^2}{6}\log y\ \mathrm{as}\ y\to \infty\ .
\end{eqnarray}

Before focusing on the physics of the system we note in passing how the thermodynamic state functions can be expanded in terms of a $y$ that can be solved for in eq. \eqref{eq:ty}. 
Everything except 
$N$ can then be expressed  by power series expansions in $y$ plus a $\log y$ times another power series expansions in $y$. This expansion can be inverted and inserted in other state functions, to give exact power series in $N$ with temperature dependence in terms of the rescaled temperature $t$.

\begin{figure}
     \centering
     \begin{subfigure}[b]{0.45\textwidth}
         \centering
         \includegraphics[width=\textwidth]{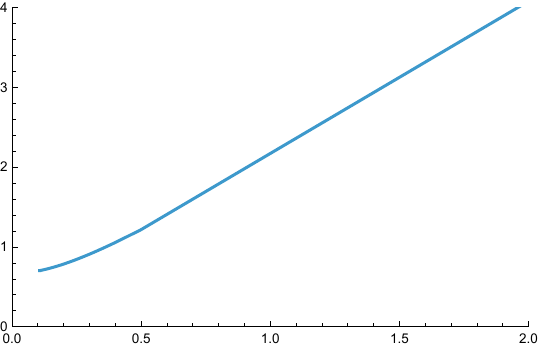}
         \caption{$rU(t)/N$}
         \label{fig:U}
     \end{subfigure}
     \hfill
     \begin{subfigure}[b]{0.45\textwidth}
         \centering
         \includegraphics[width=\textwidth]{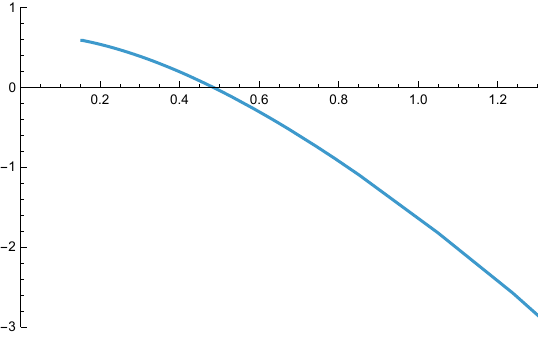}
         \caption{$rF(t)/N$}
         \label{fig:F}
     \end{subfigure}
     \hfill
        \caption{
        Thermodynamics of the boundary baryons. 
        \textbf{(a)} $U(t)$ increases almost linearly with $t$.
        \textbf{(b)} $F(t)$ vanishes for $t\approx 0.48$ and is negative for higher $t$. At the zero and at higher temperatures, the condensation of a boundary baryon is thermodynamically favoured.}
        \label{fig_baryon}
\end{figure}

By expressing the fugacity $y$ in terms of $t$ we plot dimensionless $U(t)$ and $F(t)$ with $N$ scaled out in figure \ref{fig_baryon}, based on the summary of baryon thermodynamics in table \ref{Table:StateFunctions}. The thermodynamics interpolates between a dominance of an energetic groundstate and a massless gas. We note  

\begin{table} 
\begin{center}

\begingroup
\renewcommand{\arraystretch}{1.3} 
\begin{tabular}{ |c|c|c |c| } 
 \hline
$t= r T$  & $\frac{r}{N}\,U$  & $\frac{r}{N}\,F$ & $\frac{1}{N}\,S$\\
 \hline 
  General & $-4t^3\mathrm{Li}_ 3(-y)$  & $2t^3\left\{\mathrm{Li}_ 3(-y)-\log{y}\ \mathrm{Li}_2(-y)\right\}$ & $ 2t^2\left\{ \log y\ \mathrm{Li}_ 2(-y)-3\mathrm{Li}_ 3(-y)\right\}$ \\
  $ t\to 0$  & $\frac{2}{3}$  & $\frac{4}{3}$ & $\frac{2\pi^2}{3} t  $ \\
$t\to \frac{\sqrt{6}}{\pi }$    & $\frac{18\sqrt{6}}{\pi^3}\zeta(3)$  & $ -\frac{9\sqrt{6}}{\pi^3} \zeta(3) $ & $\frac{27}{\pi^2}\zeta(3) $ \\
 $t\to \infty$    & $2t$  & $ - t\log [2e t^2]$ & $ \log [2e^3 t^2] $\\
 \hline
\end{tabular}
\endgroup

\end{center}
\caption{\label{Table:StateFunctions} Thermodynamic state functions. Denoting the length scale of the sphere, related to its light crossing time, by $R$, the distance scale $r=R/\sqrt{N}$ appears to be natural. Note that the expressions with logarithms are accurate to leading and first subleading order in the logarithms.} 
\end{table}

\begin{itemize}
\item $U$ approaches the groundstate energy $m_0\sim N^{3/2}$ for low $t$, matching the groundstate energy scaling obtained in eq. \eqref{eq:groundstate}. 
\item For high $t$ the internal energy $U \sim 2 N^{3/2} t =2N T$, as a 2d gas of $N$ relativistic particles.
\item The free energy turns negative at $t\approx 0.48$, where there is no thermodynamic barrier against the creation of a baryon, allowing for its condensation.
\end{itemize}

We should consider baryon condensation in its proper context, the full $SO(N)$ model, including arbitrary numbers of light operators, baryon operators and all other related operators. This system was studied by Shenker and Yin \cite{ShenkerYin2011} who observed a phase transition at $t_d=\frac{\sqrt{3}}{\pi}$.\footnote{The standard value for $t_d$ was obtained in the $U(N)$ model, whereas we wish to compare to a baryon condensation temperature in the $SO(N)$ model. In appendix \ref{app_deconf} we check that $t_d$ is unaffected by the change of group to leading order.} For a very brief reminder, see appendix \ref{app_thermo}. We find that a baryon condenses at a \emph{lower} temperature $t_c$ than the deconfinement transition, $t_c<t_d$, with
\begin{equation}
F(t_c)=0 ,
\end{equation}
as seen in figure \ref{fig:F}.

Recall from section \ref{HeavyOperators} that there are no proper two-baryon operators. This means that the condensation of a baryon is well-behaved. When it is statistically favourable to create a single baryon, creation of two or more baryons, if allowed, could also be favourable above $t_c$, leading to a more dramatic behaviour, probably a phase transition. The absence of multi-baryon states may thus explain why there is no dramatic feature in the free energy of the full $SO(N)$ model below $t_d$. There are, however, signals of other interesting features in the system below the critical temperature, detectable by inelastic scattering \cite{Engelsoy:2021fbk}.

\section{Dual bulk objects and their thermodynamics}\label{sect_DualObjects}

\subsection{Dual bulk objects}

We would like to understand if any bulk objects behave thermodynamically as duals to the heavy boundary operators. Bulk duals to thermal ensembles in the boundary are associated to black holes (or to thermal AdS at low $T$). In the present case, however, we do not consider the dual of the full theory, but a thermal ensemble of single baryon states. We then expect a thermal ensemble of different excitations of a single bulk object rather than an ensemble of bulk multiparticle states. Already the description of baryon states as massless gas states on $S^2$ suggests that this single bulk object is membrane-like with an $S^2$ topology 
--- just as massless gases on $S^1$ are associated to excitations of closed strings.

Since there are no other parameters than $R$ and $N$ in the baryon spectrum, we might ask if there are candidate membrane-like objects which are equally parameter free. Fundamental tensionless membranes are such candidates, but it is unclear why fundamental membranes would have a high threshold energy $E_0 \sim \frac{2N^{3/2}}{3R}$, from eq.\ \eqref{eq:groundstate}, inversely proportional to a power of Newton's constant $G_N\sim \frac{L^2}{N}$ of vector model holography. Here, $L$ is the AdS length scale. The threshold energy instead indicates that the correct membrane system includes gravitational self-interactions just as black holes do.

Objects called `black shells' have been proposed as black hole mimickers, ie horizonless compact objects which can replace black holes. Black shells are idealised objects with an infinitely thin shell separating a bubble of spacetime with a distinct cosmological constant from the surrounding vacuum. Ultimately, the appearance of the system from the outside only depends on the mass or temperature of the black shell, Newton's constant and the cosmological constant of the outside. There is no dependence on a shell tension or the cosmological constant in the interior. These properties make black shells promising candidates for bulk duals of baryons.

We refer to the literature, eg.\ \cite{Danielsson:2017riq,Danielsson:2025vyd}, for details about black shell physics and its underlying assumptions, which we only sketch here. The dynamics of the inner vacuum bubble, the thin shell, and the external AdS geometry is governed by standard junction conditions, and the nature of the stress tensor of the thin shell. An effective perfect fluid stress tensor distinguishes these shells from other objects, and makes them suitable for the present purpose. The self-interacting gravity/shell system can be understood either through  the local stress tensors on the shell or the total energies measured at infinity. 
After compensating for vacuum energy and pressure terms from the interior bubble vacuum, the equation of state of black shells is assumed to be that of radiation on the shell.\footnote{In our context, massless equations of motion on the shell would be due to conformal coupling in the boundary theory. In other contexts, the effective massless equations have been motivated by D-brane considerations \cite{Danielsson:2017riq}.} Static solutions may then be found.

Fortunately, the thermodynamics of black shells was recently studied in AdS \cite{Danielsson:2025vyd}, enabling us to take high temperature limits of known expressions to compare with boundary baryons. Since black shells are defined as defects in Einstein gravity, rather than in higher spin gravity, the conjectured dual of the vector model, a perfect match is not to be expected. Nevertheless, at high $T$, the reasonable correspondence limit, we will find that the functional dependence and the asymptotics match perfectly upon fixing two entries in the holographic dictionary. 

\subsection{Bulk black shell thermodynamics}

Embedding spherically symmetric black shells in Einstein gravity permits standard junction conditions to relate the shell energy density, the shell pressure and the parameters of the metric. The resulting thermodynamic state functions in an AdS background can be found in ref. \cite{Danielsson:2025vyd} as function of the area of the spherical shell, or the related radial coordinate $r_0$, here expressed in terms of a geometric dimensionless measure
\begin{equation}
 x=r_0/L
\end{equation}
of the shell size in terms of the AdS scale.

The expressions
\begin{eqnarray}
\frac{GM_{s}}{L}&=& \frac{4 x}{9}\frac{1+\frac{3 x^2}{2 }}{1+x^2},\\
L T_{s}&=& \frac{x}{2 \pi }\left(1+\frac{4 }{9x^2}\frac{1+\frac{3 x^2}{2}}{1+x^2}\right),\\
\frac{G F_{s}}{L}&=& \frac{4 x}{9}
\frac{1+\frac{3 x^2}{2 }}{1+x^2}
-x\left(1+\frac{4 }{9x^2}\frac{1+\frac{3 x^2}{2 }}{1+x^2}\right)
\left[  \log \left(1+x^2\right)-\frac{2}{3} \log \left(1+\frac{3 x^2}{4}\right)\right],\\
\frac{G S_{s}}{L^2}&=&2\pi \log (1+x^2)-\frac{4\pi}{3}  \log (1+\frac{3 x^2}{4}),
\end{eqnarray}
for shell quantities show that Newton's constant $G$ simply provides a conversion factor of geometry into thermodynamically meaningful quantities. Black shells are very simple objects and do not depend on any free parameter which could combine with $G$ in a nontrivial way.
As in the case of the baryon, thermodynamics is expressed in terms of a dimensionless parameter which should be eliminated to give the direct thermodynamic relations. Here, $x$ has no obvious boundary meaning, while in the boundary calculation, the fugacity $y$ lacks an obvious bulk meaning. 

\begin{figure}
     \centering
     \begin{subfigure}[b]{0.45\textwidth}
         \centering
         \includegraphics[width=\textwidth]{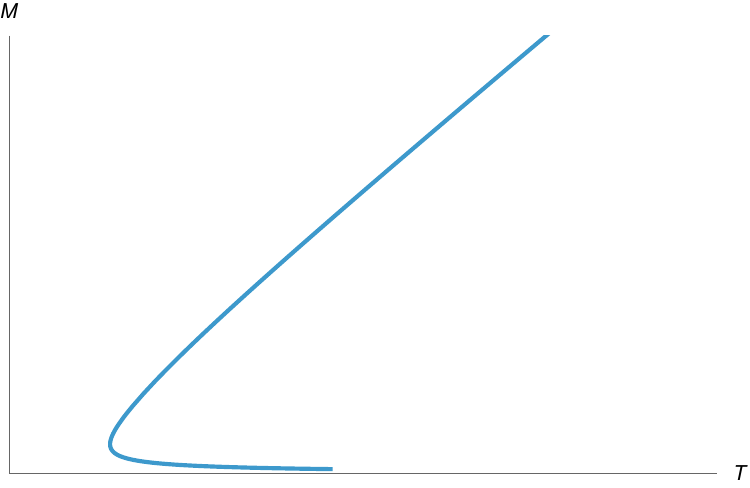}
         \caption{$M_s(T)$}
         \label{fig:U}
     \end{subfigure}
     \hfill
     \begin{subfigure}[b]{0.45\textwidth}
         \centering
         \includegraphics[width=\textwidth]{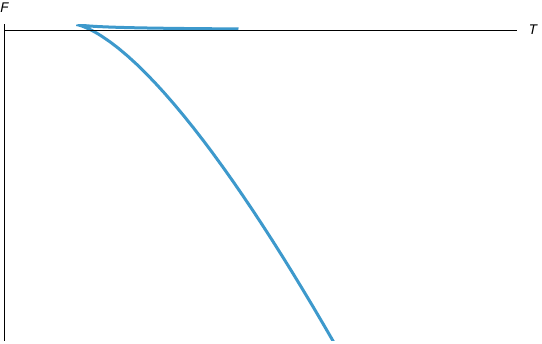}
         \caption{$F_s(T)$}
         \label{fig:F}
     \end{subfigure}
     \hfill
        \caption{Thermodynamics of black shells. Both plots display one large black shell branch and one small black shell branch, analogous to small and large black holes in AdS and originating at a common solution with minimal temperature. 
        \textbf{(a)} 
The upper, large black shell branch has a minimal $M_\text{min}$. 
        \textbf{(b)} 
        The large black shell branch has negative free energy above a condensation temperature. 
        }
        \label{fig_BlackShell}
\end{figure}

Black shell thermodynamics is summarised in figure \ref{fig_BlackShell} and table \ref{Tab_BS}, analogously to the descriptions of boundary baryons in section \ref{sec_BoundaryBaryon}. 
It is clear from the plots that the low temperature behaviour of boundary baryons and black shells is quite different, although black shells are also likely to condense at a certain temperature \cite{Danielsson:2025vyd}. Differences are expected because the bulk theory dual to free vector models is different from Einstein gravity. The question is whether there are any quantitative similarities at all is answered in the next section.

\begin{table} 
\begin{center}

\begingroup
\renewcommand{\arraystretch}{1.5} 
\begin{tabular}{ |c|c|c|c| } 
 \hline
  $t_s=LT$  & $\frac{G}{L}\,M_s$  & $\frac{G}{L}\,F_s$ & $\frac{G}{L^2}\,S_s$\\
\hline 
  $\frac{x}{2 \pi }\left[1+\frac{4 }{9x^2}\frac{1+\frac{3 x^2}{2}}{1+x^2}\right]$  & $\frac{4x}{9}\frac{1+\frac{3x^2}{2}}{1+x^2}$  & $\frac{G}{L}(M-TS)$ & $2\pi \log (1+x^2)-\frac{4\pi}{3}  \log (1+\frac{3 x^2}{4})$ \\
$t_s \to \frac{7}{9 \pi }\approx 0.25$    & $\frac{5}{9}$  & $\frac{5}{9}-\frac{14}{27} \log \frac{128}{49}\approx 0.058 $ & $\frac{2}{3} \pi  \log \frac{128}{49}$ \\
 $t_s\to \infty$    & $\frac{4\pi}{3}t_s$  & $ -\frac{2\pi}{3}t_s\log{[\frac{64 \pi^2}{9e^2}t_s^2]}$ & $\frac{2\pi}{3}\log{[\frac{64}{9} \pi^2 t_s^2]}$\\
\hline
\end{tabular} 
\endgroup

\end{center}
\caption{\label{Tab_BS} Black shell state functions in different limits of $T$. $L$ is the AdS length scale, the return time in pure $AdS$, which is also the return time for signals far from the central mass. The logarithmic terms are accurate to leading and first subleading order in the logarithms.
}
\end{table}

\section{Discussion of baryon-black shell match}\label{sect_Discussion}

At higher and higher temperatures both black holes and black shells extend over larger and larger distances, much longer than the AdS length $L$. As the physical length scale increases, we expect less dependence on the high energy states of the theory. Since the main difference between Einstein gravity and higher spin theory in AdS is its high energy spectrum, it is worthwhile to check the high temperature limits of black shells and boundary baryons.\footnote{In particular, this argument is relevant in global AdS.} In this correspondence limit, there is a chance of a match.

Inspection of the high temperature limits of baryons (table \ref{Table:StateFunctions}) and black shells (table \ref{Tab_BS}) reveals similar behaviour. Equating the black shell mass and the energy of the boundary baryon yields
\begin{equation}
\frac{4\pi}{3}\frac{L^2}{G}T=2NT,\quad \text{ie} \quad \frac{1}{N}=\frac{3G}{2\pi L^2},
\end{equation}
which also matches the leading logarithmic terms in the entropy and the free energy. This is the standard identification of  large $N$ in weakly coupled vector model holography, where the precise numerical factors depend on the conventions in the definition of $L$. Thus, we have established that high temperature baryons behave thermodynamically as black shells.

Encouraged by this success, we may ask what matching of the next-to-leading terms in the free energy and entropy would lead to. The corresponding condition is 
\begin{equation}
2e r^2 T^2 = \frac{64 \pi^2}{9e^2}L^2T^2,\quad \text{ie} \quad R^2 =N\frac{32 \pi^2}{9e^3}L^2\ ,
\end{equation}
which relates bulk and boundary length scales, surprisingly in an $N$ dependent way. To our knowledge, such a relation is neither established, nor in conflict with any firmly established principles. One can regard this relationship as a determination of a tree level cosmological constant $\Lambda \sim \frac{1}{R^2}N$.\footnote{Standard determination of the cosmological constant has been hampered by the lack of an ordinary bulk action for higher spin theory, but see ref.\ \cite{Aharony:2022feg} for an evaluation in terms of a bilocal action which yields the same scaling with $N$ as the argument presented here.}  If we use this relationship for order of magnitude estimates, despite the unreliability of the duality at low temperatures, we find that a typical low black shell mass
$$
M_s \sim \frac{L}{G} \sim\frac{N^{3/2}}{R}
$$
which is of the same order as a typical low baryon mass, cf. the groundstate mass $m_0=E_0$ from eq.\ \eqref{eq:groundstate}.

\section{Conclusion}\label{sect_Conclusion}

This study originated in the general puzzle of heavy operators in conformal field theory. They appear to be simultaneously disposable for standard low energy observables and crucial for both the internal consistency and for holographic duals of bulk black holes. Free vector models are among the simplest examples available for their study, and the benchmark study by Shenker and Yin \cite{ShenkerYin2011} raises as many interesting questions as it answers.

In summary, the main hard findings of the present work on vector models are: concrete thermodynamic descriptions of baryons in the free 3d $SO(N)$ models on the sphere; including a condensation temperature $T_c \approx 0.87 T_d$, lower than the deconfinement temperature $T_d$; and its high temperature limit as a relativistic gas of $N$ particles. In addition, the high temperature limiting expressions for black shell thermodynamics in AdS$_4$ were obtained and found to agree with the baryon results to leading order, and to slightly higher accuracy upon relating bulk and boundary distance scales as if the AdS cosmological constant scales as $N$ for a fixed boundary radius. We have also argued that a much better match is not to be expected. In particular, since we have not considered supersymmetric systems, they are not heavily constrained. 

There are still puzzles awaiting explanation, both internal to the $SO(N)$ model and concerning holographic duality. As mentioned in section \ref{HeavyOperators}, many other secondary invariant operators similar to baryons are expected in vector models but their spectrum has as yet only been studied carefully in vector quantum mechanics models \cite{deMelloKoch:2025cec}, not in quantum field theory. These operators should contribute to the general thermodynamics at temperatures around condensation and higher --- thus also to the deconfinement transition, and eventually to microstates of large bulk black holes. It is in fact reasonable to regard the Fermi gas as a template for more general secondary invariants in vector models.\footnote{I thank Robert de Mello Koch for this comment.}

The nature of black holes in duals of vector models and, possibly equivalently, in Vasiliev's higher spin theory, has been debated \cite{ShenkerYin2011,AmadoSundborgThorlaciusWintergerst2017,AmadoSundborgThorlaciusWintergerst2018,DidenkoVasiliev2009,Iazeolla:2017vng}. For intriguing new ideas, see refs. \cite{David:2020fea,Neiman:2022enh,Lang:2024dkt}. Our present findings add another ingredient to the puzzle: what is the role of simple heavy states in the bulk? Do black shells and their relatives, the secondary invariants, span an over-complete space of configurations which saturate a black hole Hilbert space?

Generally, we would like simple vector models, matrix models and their invariants to illuminate the structure of black hole Hilbert spaces. The present study shows that there may surprises along that route.

\appendix
\section{Thermodynamics in the $U(N)$ model}\label{app_thermo}

The thermodynamics of \emph{all} states, including multiparticle excitations of light states, is captured by the canonical partition function of the vector model, restricted to physical (gauge invariant) states. The computation \cite{ShenkerYin2011} replaces characters for adjoint matter in \cite{Sundborg2000} by characters of vector matter. Writing
\begin{equation}
Z_{U(N)}= \prod_{m=1}^N \int_{-\pi}^\pi d\alpha_m e^{ S_{U(N)}}
\end{equation}
where the integral comes from the projection to singlet (gauge invariant) states. One finds
\begin{equation}
S_{U(N)}=\sum_{1\le i<j\le N} V_2(\alpha_i,\alpha_j) + \sum_{i=1}^N V_1(\alpha_i)
\end{equation}
with two--and one-eigenvalue potentials
\begin{eqnarray}
V_2(\alpha_i,\alpha_j) &\equiv &2 \ln \left|\sin \left(\frac{\alpha_i-\alpha_j}{2}\right)\right| \\
V_1(\alpha_i) &\equiv & 2 N_f \sum_{m=1}^{\infty} \frac{1}{m} z_S\left(x^m\right) \cos \left(m \alpha_i\right)
\end{eqnarray}
%
from the group invariant Haar measure, here on $U(N)$, and the matter content, through the partition function $z_S\left(x^m\right)$ of the constituent field and the character of the matter representation (here fundamental + antifundamental of $U(N)$).

Saddle points of the effective action $S_{U(N)}$ for a charge distribution on the circle dominates the integral for large $N$ and yield a transition temperature $T_d=\frac{1}{\pi}\sqrt{3N}$, where the eigenvalue distribution of the saddle develops a gap. 
Intuitively, one would expect general large $N$ properties and not the type of group, $U(N)$ or $SO(N)$ to control the leading behaviour, as asserted in \cite{ShenkerYin2011}. Since this is important for our physical conclusions we confirm this expectation explicitly in the next appendix.

\section{The deconfinement temperature in the $SO(N)$ model}\label{app_deconf}

In short, our argument compares the eigenvalue effective action for $SO(N)$ with the effective action of $U(N)$ evaluated for eigenvalue configurations symmetric under reflections $\alpha \to -\alpha$  (to which the saddle configuration belongs). The balance of terms which determines $T_d$ is not upset to leading order in $N$, which implies that $T_d$ is unchanged by the change of large rank classical group, to leading order.

The reflection symmetric saddles are then given by labelled pairs of eigenvalues $(-\alpha_i,\alpha_i)$, where each $\alpha_i \in [0,\pi]$. The action becomes
\begin{equation}
S^{sym}_{U(2k)}=2\left\{\sum_{1\le i<j\le k}V_2(\alpha_i,\alpha_j)+\sum_{1\le i <j\le k } V_2(\alpha_i,-\alpha_j)
+\frac{1}{2}\sum_{i=1}^{k} V_2(\alpha_i,-\alpha_i)  +\sum_{i=1}^{k} V_1( \alpha_i)\right\}
\end{equation}
for even $N=2k$.
Comparing with the action obtained from the Haar measure on $SO(2k)$ and matter in the fundamental (vector) representation of $SO(2k)$,
\begin{equation}
S_{SO(2k)}=
\sum_{1\le i<j\le k}  \left( V_2(\alpha_i,\alpha_j)+V_2(\alpha_i,-\alpha_j) \right)
+ \sum_{i=1}^{k} V_1( \alpha_i)
\end{equation}
we find
\begin{equation}\label{eq:SO(2k)}
S_{SO(2k)}=
\frac{1}{2}S^{sym}_{U(2k)} -\sum_{i=1}^{k} V_2(\alpha_i,-\alpha_i) 
\end{equation}
provided the last term can be disregarded. 
Since the last term only involves a single sum, it can be neglected in the large $N$ limit\footnote{It corresponds to repulsive interactions of the other eigenvalues with fixed `trivial' eigenvalues at $\alpha=0$ and $\alpha=\pi$. It may be regarded as a temperature independent correction to $z_S$.}. We conclude that the saddle point equations are the same for $SO(2k)$ with vector matter as for the reflection symmetric sector of $U(2k)$ with fundamental plus antifundamental matter, in the large $N$ limit. The critical temperature is thus also the same to leading order. A similar argument applies to $SO(2k+1)$.

\acknowledgments
I would like to thank Anders K H Bengtsson, Ulf Danielsson, Slava Didenko, Oscar Henriksson, Julian Lang, Andrei Parnachev, Sanjaye Ramgoolam, Per Sundell, and Larus Thorlacius for conversations related to this work, and especially Robert de Mello Koch and Zhenya Skvortsov who also made valuable comments on the manuscript. Part of the work was done at the Mons Higher Spin meeting and part at the Simons Centre programme on black holes.

\bibliography{baryon}
\bibliographystyle{JHEP}

\end{document}